\documentclass[showpacs, pra,twocolumn,preprintnumbers ,amsmath, amssymb, superscriptaddress, aps]{revtex4-2}

\usepackage{color}
\usepackage{amsmath,amssymb}
\usepackage{pifont}% http://ctan.org/pkg/pifont
\usepackage{amssymb}  % More symbols
\usepackage{bbold}
\usepackage{float}
\usepackage{subfloat}

\usepackage[caption=false]{subfig}
\usepackage{tikz}
\usepackage{makecell}
\usepackage{subfig}
\usepackage{pifont}   % Ding symbols
\usepackage{graphicx} % Include figure files
\graphicspath{{Figures/}}
\usepackage{dcolumn}  % Align table columns on decimal point
\usepackage{bm}       % bold math
\usepackage{multirow} % Table functions
\usepackage{placeins}% For making floats not move around everywhere
\usepackage[colorlinks]{hyperref}
\usepackage{mathtools}
\usepackage{appendix}

\def \be{\begin{align}}
	\def \ee{\end{align}}
\def \bea{\begin{eqnarray}}
	\def \eea{\end{eqnarray}}

\begin{document}
	
	\title{Laser-induced modulation of conductance in graphene with magnetic barriers}
		
		%Laser-controlled conductance in graphene with magnetic barrier structure}
		
		%Conductance through double magnetic barriers assisted by a laser field in graphene}
	
	\author{Rachid El Aitouni}
	\affiliation{Laboratory of Theoretical Physics, Faculty of Sciences, Choua\"ib Doukkali University, PO Box 20, 24000 El Jadida, Morocco}
	\author{Miloud Mekkaoui}
	\affiliation{Laboratory of Theoretical Physics, Faculty of Sciences, Choua\"ib Doukkali University, PO Box 20, 24000 El Jadida, Morocco}

\author{Pablo Díaz}
\affiliation{Departamento de Ciencias F\'{i}sicas, Universidad de La Frontera, Casilla 54-D, Temuco 4811230, Chile}  
\author{David Laroze}
\affiliation{Instituto de Alta Investigación, Universidad de Tarapacá, Casilla 7D, Arica, Chile}
	\author{Ahmed Jellal}
	\email{a.jellal@ucd.ac.ma}
	\affiliation{Laboratory of Theoretical Physics, Faculty of Sciences, Choua\"ib Doukkali University, PO Box 20, 24000 El Jadida, Morocco}
%	\affiliation{Canadian Quantum  Research Center,
%		204-3002 32 Ave Vernon,  BC V1T 2L7,  Canada}
	\begin{abstract}

We study how electrons move across a graphene sheet when it encounters two magnetic barriers with a region in between that is continuously driven by laser light.
Rather than acting as a static obstacle, this illuminated middle section becomes a Floquet cavity that opens new transport channels through controlled photon absorption and emission. By combining Floquet theory with the transfer matrix method, we track electron transmission through both the main energy band and the emerging photon-assisted sidebands.
We find that the laser does more than modify the potential—it reshapes how electrons interact between the magnetic barriers, enabling a switch from ordinary transmission to transport dominated by photon exchange. Because the magnetic field and the optical drive are applied to separate sections of the device, the system supports interference between cyclotron-filtered motion and discrete photon-pumping channels, producing Fano resonances and angle-dependent transmission zeros that cannot appear in double magnetic or double laser barrier systems alone.
Under well-defined conditions, the distance between the magnetic barriers controls the coupling between Floquet channels, allowing highly tunable resonances and even perfect transmission, despite strong magnetic confinement. We also observe that low-energy carriers are efficiently blocked by the magnetic regions, while conductance steadily rises with energy until it reaches a clear saturation plateau.
This hybrid design provides a versatile way to steer graphene electrons by balancing optical pumping and magnetic momentum filtering.

	\end{abstract}

	\maketitle

\section{Introduction}
Graphene is a two-dimensional material consisting of a single type of carbon atoms arranged in a honeycomb structure. It was isolated for the first time in 2004 \cite{Novos2004} and later it is demonstrated that it is rigid, flexible \cite{prop} and can be transparent, only $2.3\%$ of light is observed \cite{absor}. Within the framework of the tight-binding model, the behavior of electrons in graphene is analyzed \cite{Tight}. These electrons exhibit properties similar to massless charged particles, moving at a speed 300 times slower than the speed of light \cite{mobil,mobil2}. For this reason, they are often referred to as massless Dirac fermions \cite{masless}. The presence of two atoms in the unit cell appears in the energy spectrum as two non-equivalent points $K$ and $K^\prime$, called Dirac points. The energy spectrum near these points is conical, with the two bands touching at these points. However, there is no gap between the conduction band and the valence band \cite{zero,zero1}. The absence of a gap between the two bands is the main problem with graphene, as all the fermions move freely between these bands and cannot be controlled. Despite all the incredible properties of graphene, its use in technology remains limited due to the uncontrollability of these fermions. Creating this gap paves the way for the use of graphene in the fabrication of electronic components, such as field-effect transistors, sensors, photodetectors \cite{Novos2004, Geim2007, Schwierz2010}.

 Several research efforts have been made to create a band between the two bands to control the motion of fermions in graphene. Among these methods, deformation of graphene, which generates a pseudomagnetic field, allows the creation of a band gap \cite{def1,def2}. In addition, placing graphene on a substrate can also create a band gap \cite{substrat}. Doping graphene with a different type of atom allows for variation of the energy spectrum and the creation of a band gap between the valence and conduction bands \cite{dopage}. Applying a simple rectangular barrier can create a gap \cite{klienexp}, but another problem that arises is total transmission for normal incidence, known as the Klein tunneling effect \cite{klien1,klien2}. In this case, the transmission is independent of the width and height of the barrier. The use of time-oscillating barriers allows the energy spectrum to degenerate, resulting in multiple transmission modes, each corresponding to energy sub-bands \cite{oscil1,oscil2,timepot,timepot2,doubletemps}.
  This results in two transmission processes: transmission with photon exchange between the barrier and the fermions corresponding to the side bands, and the process without photon exchange corresponding to the central band. The majority of fermions crossing the barrier do so without photon exchange, and the probability for the more distant side bands remains low. The magnetic barrier is also a tool to create the forbidden band, with the degeneration of the energy spectrum by the appearance of Landau levels \cite{mag1,mag3,mag4,magneticfield}. The confinement of fermions remains limited due to the presence of the Klein tunneling problem. Combining a magnetic barrier with a time-oscillating potential works better than using simple barriers alone, although the confinement is still partial \cite{bis2}. Laser irradiation is also used for fermion confinement, creating sub-bands that result in an infinite number of transmission modes. However, the majority of fermions cross the barrier through the central band. The intensity of the laser irradiation can suppress the Klein tunneling effect and confine fermions at certain levels \cite{Elaitouni2023}. Double magnetic barriers can also confine fermions \cite{doublemag}, as can double laser barriers \cite{doublelaser,doublelaser+magn}, but the confinement is not perfect.

  We propose a confinement method based on the combination of laser and magnetic fields. Specifically, our confinement consists of double magnetic barriers separated by an intermediate region irradiated by a monochromatic linearly polarized laser field. The magnetic barriers are formed by nanostructured ferromagnetic strips placed over the graphene layer. %The average magnetic field for both barriers is zero
  Since the proposed fields are oriented in opposite directions, one upward and the other downward, the resulting average magnetic field across both barriers is zero \cite{conduct2}.  We show that a time-oscillating laser field quantizes the energy spectrum, leading to two distinct transmission processes: one with photon exchange and one without. It is found that the transmission is blocked by the magnetic field and the laser irradiation. The perfect transmission remains and is observed in a left-shifted region, while the anti-resonances (perfect reflection) are observed in a small region.
  Increasing the distance between the barriers decreases the transmission, indicating a reduction in the number of fermions crossing the barrier. This leads to a decrease in the conductance, which is consistent with observations for double magnetic and double laser barriers, since increasing this distance increases the probability that fermions undergo destructive interference before crossing the second barrier.
  It is also noted that the tunneling effect for our system is significantly different from that of a double magnetic barrier \cite{doublemag}, a double laser barrier \cite{doublelaser}, and a typical double electrostatic barrier. 
  {The tunneling difference originates from the simultaneous energy modulation and spatial separation of magnetic and laser regions, which introduces interference between cyclotron momentum filtering (magnetic barriers) and discrete Floquet sideband channels (laser region). Since photons are exchanged only in the region between the magnetic barriers, the transmission exhibits field-mediated channel coupling governed by boundary matching conditions, rather than static barrier tunneling alone. This leads to a qualitative change in resonance structure and carrier confinement, which is absent when only magnetic or laser fields are applied individually.}

The paper is organized as follows. In Sec. \ref{TM}, we present the theoretical model describing our system and derive the solutions for the energy spectrum in each region separately. In Sec. \ref{cond}, we apply the boundary conditions together with the matrix transfer approach to explicitly determine the transmission and conductance. To provide a deeper understanding and illustration of our results, we perform numerical analysis in Sec. \ref{TP}, providing various discussions and comparisons with existing literature. We conclude the manuscript in Sec. \ref{CC}.

To analyze these phenomena quantitatively, we now introduce the theoretical framework that describes electron transport in graphene under combined magnetic and laser barriers. This model lays the foundation for understanding the transmission channels, Floquet sidebands, and the mechanisms leading to Fano resonances and anti-resonances.

\section{THEORETICAL MODEl}\label{TM}

\begin{figure}[H]
	\centering
	\includegraphics[scale=0.13]{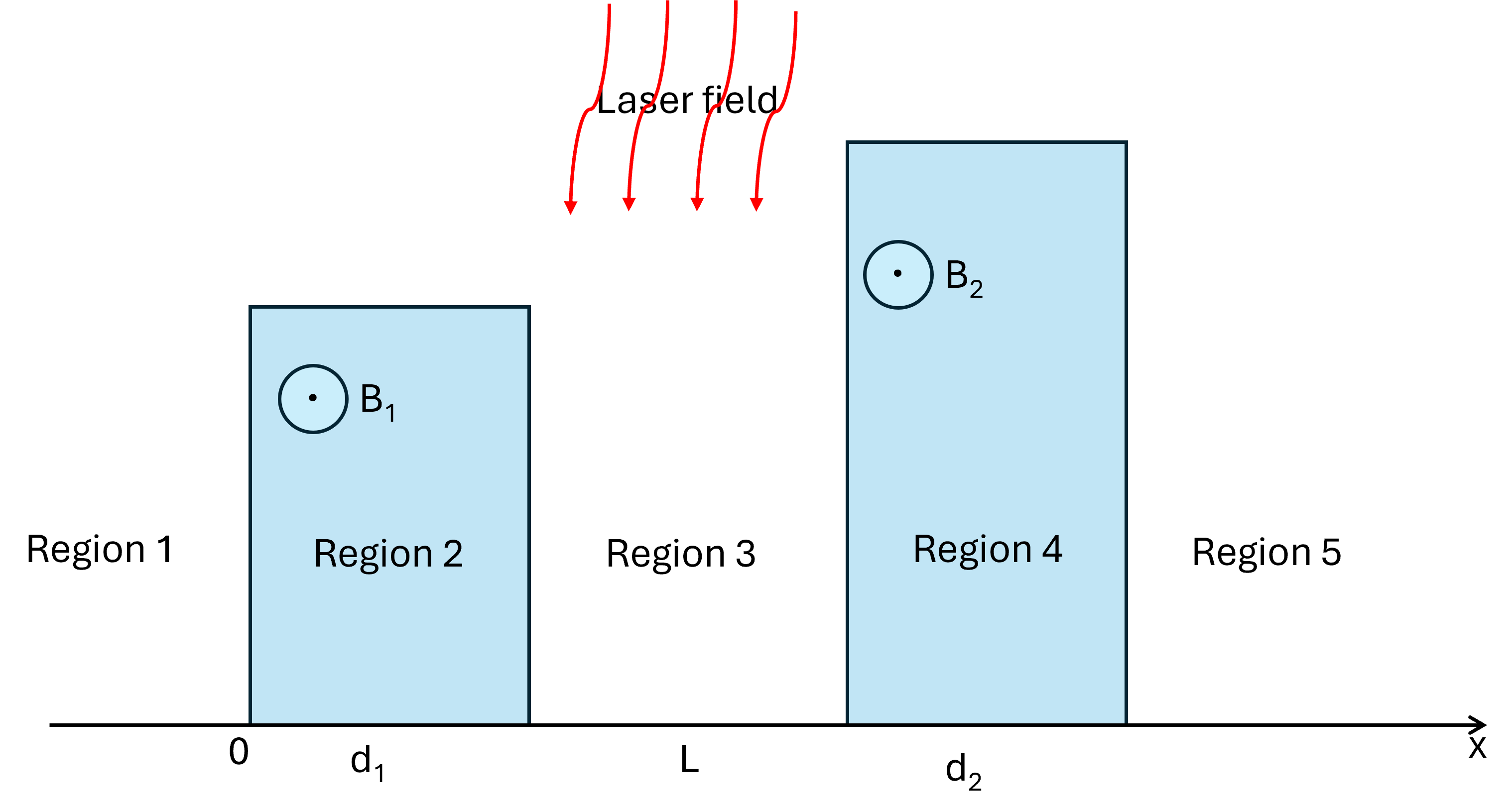}
	\caption{Schematic of a graphene sheet divided into 5 regions, where regions 1 and 3 are controlled by two magnetic fields of different amplitudes and region 2 is irradiated by a linearly polarized laser field.}\label{fig11}
\end{figure}

We apply two different magnetic fields, $B_1$ and $B_2$, to a graphene sheet to study the resulting effects on fermion dynamics. The first magnetic field, $B_1$, is applied to region 1, which has a width denoted as $d_1$. The second magnetic field, $B_2$, is applied to region 3, which has a width denoted $d_2$. These two magnetically influenced regions are separated by an intermediate region of width $L$. This intermediate region is controlled by a laser field generated by a periodic electric field. 
Two practical configurations generate static magnetic barriers in graphene. The first uses two ferromagnetic bars placed in opposite directions, producing a stable magnetic region with sharp interfaces \cite{mag1,conduct2}. The second approach employs lithographically patterned electrodes that generate a localized magnetic profile via controlled currents \cite{mag3}. While we model the barriers using a $\delta$-function approximation, finite-width implementations capture the same qualitative effects on electron transmission.
For a better illustration, we visualize the system profile as shown in Fig. \ref{fig11}. The positions of the barrier interfaces are identified as $x_1=x_2=0$, $x_3=d_1$, $x_4=d_1+L$ and $x_5=d_1+L+d_2$.

Mathematically, we introduce the vector potential $\boldsymbol{A}_{L}$, which corresponds to the laser field generated by a sinusoidal electric field $\boldsymbol{E}_{L} = -\frac{\partial \boldsymbol{A}_{L}}{\partial t}$ in the dipole approximation \cite{dipole}. Here, we assume that the laser wavelength is much larger than the lattice constant and that its intensity is low enough for the electronic response to remain in the linear regime \cite{aprox}. They are given by
\begin{align}
\boldsymbol{A}_{L}=(A_0\cos(\omega t), 0), \quad 	\boldsymbol{E}_{L}=(E_0 \sin(\omega t), 0)
\end{align}
with relation $E_0=A_0\omega$. We also consider two barriers created by the following magnetic fields applied along the $z$-direction
 \begin{align}
&\boldsymbol{B}_1=B_1\left[\delta(x)-\delta(x-d_1)\right]\boldsymbol{e}_z\\
& \boldsymbol{B}_2=B_2\left[\delta(x-(d_1+L))-\delta(x-(d_1+L+d_2))\right]\boldsymbol{e}_z.	
 \end{align} 
These fields give rise to  the  vector potential $\boldsymbol{A}_B$  in the Landau gauge for the five regions as
 \begin{align}	A_B(x)=
	\begin{cases}
		0,&x<0\\
		A_{1}=B_1 l_B,&0<x<d_1\\
		0,&d_1<x<L+d_1\\
		A_2= B_2 l_B,&d_1+L<x<d_1+L+d_2\\
		0,&x>d_1+L+d_2
	\end{cases}
\end{align} 
and we have included the magnetic length $l_{B}=\sqrt{\frac{1}{B}}$ in the unit system ($\hbar=e=c=1$), where $B_1$ and $B_2$ are the field intensities. For the case of a single magnetic barrier, see \cite{conduct2}.

The present system can be described by an appropriate Hamiltonian near the Dirac point $K$, where the low-energy excitations in graphene behave as massless Dirac fermions. The effective Hamiltonian captures the dynamics of these fermions as they are affected by the external fields of the system. Near the Dirac point, the Hamiltonian takes the form
\begin{equation}\label{Ham}
H=	v_{\mathrm{F}} \boldsymbol{\sigma} \cdot\left( \boldsymbol{p}+\frac{e}{c} \boldsymbol{A}_B+\frac{e}{c} \boldsymbol{A}_L\right) %=E \psi
\end{equation}
where $v_{F}$ is the Fermi velocity, $\boldsymbol{\sigma}=\left(\vec{\sigma}_{x}, \vec{\sigma}_{y}\right)$  are the Pauli matrices and $\boldsymbol{p} $ is the momentum.
Since the oscillating laser field creates multiple energy bands, the solution to the time-dependent eigenvalue equation $H\psi_{j}(x,y,t)=E\psi_{j}(x,y,t)$
can be expressed as a sum of Floquet modes  $m\omega$ ($m\in\cal{Z}$), where the wavefunction is expanded as
 \cite{oscil1,oscil2,timepot,timepot2,doubletemps}
\begin{equation}\label{Fwave}
	\psi_{j}(x,y,t)=\phi_j(x,y)\sum_{m=-\infty}^{\infty}J_m(\alpha)e^{-iv_F(\varepsilon+m\varpi) t}
\end{equation}
where we set $\varpi=\frac{\omega}{v_F}$, $\varepsilon=\frac{E}{v_F}$, and the parameter $\alpha=\frac{E_0}{\varpi^2}$ which characterizes the laser field. Here we used the generating function
of the Bessel function $J_m(\alpha)$, which is equal to
\begin{equation}
e^{i \alpha \sin(\varpi t)} = \sum_{m=-\infty}^{\infty} J_m(\alpha) e^{-i m \varpi t}.	
\end{equation}
Note that the wave vector component \( k_y \) is a conserved quantity, then we can express the spinor associated with $H$ as \( \phi_{j}(x,y) = \begin{pmatrix} \phi_{j}^A(x) \\ \phi_{j}^B(x) \end{pmatrix} e^{i k_y y} \). Furthermore, due to energy conservation, the two pseudospin states of the fermions in the \( j \)-th region can be expressed as a linear combination of wave functions with energies \( \varepsilon_j + m\varpi \). Solving the eigenvalue equation introduces a new index \( l \in \mathbb{Z} \) corresponding to the Floquet sideband energy states \( \varepsilon_j + l\varpi \). Thus, we can express the solution as
\begin{equation}
	\psi_{j}(x,y,t)=e^{i k_{y} y}\sum_{m,l=-\infty}^{\infty}\phi^{(l)}_{j}(x,y)J_{m-l}(\alpha_j)e^{-iv_F(\varepsilon+\varpi)t}.
\end{equation}
To proceed and find the solutions of the energy spectrum, we need to solve the Dirac equation in each region independently. This involves taking into account the external fields applied to each region.

In regions 1 and 5, there is only pristine graphene, and therefore  the corresponding spinors can be easily expressed as \cite{Elaitouni2022,Elaitouni2023}
\begin{widetext}
\begin{align}
&\psi_{1}(x,y,t)=e^{i k_{y} y} \sum_{m,l=-\infty}^{\infty}\left[
\begin{pmatrix}
1\\	\gamma_l
\end{pmatrix}
e^{i k_l(x-x_0)}\delta_{m,0}+r_l 
\begin{pmatrix}
	1\\-\frac{1}{\gamma_l}
\end{pmatrix}
e^{-i k_l (x-x_0)}\right]\delta_{m,l}e^{-iv_F(\varepsilon+m\varpi) t} \tag{4}\\
&\psi_{5}(x,y,t)= e^{i k_{y} y}\sum_{m,l=-\infty}^{\infty}t_l 
\begin{pmatrix}
	1\\ \gamma_l
\end{pmatrix}
e^{i k_l (x-x_4)}\delta_{m,l}e^{-iv_F(\varepsilon+m\varpi) t} \tag{4}\\
& \gamma_l=\frac{k_l+i k_{y}}{\varepsilon+l\varpi}
\end{align}
\end{widetext}
related to the energies
\begin{align}\label{E8}
	\varepsilon+l\varpi=s_l\sqrt{k_l^2+k_y^2}
\end{align} 
where we  defined  $	\delta_{m,l}=J_{m-l}(0)$ and 
$s_l=\text{sign}(\varepsilon+l\varpi)$. From \eqref{E8}, we can derive the propagating modes along $x$-direction as
\begin{align}
	k_l=\sqrt{(\varepsilon+l\varpi)^2-k_y^2}.
\end{align}

In regions 2 and 4, a static magnetic field is present. From the eigenvalue equations, we obtain two coupled equations as follows
	\begin{align}
	&	\left[-i\partial_x-i(k_y+A_B)\right]\phi_{j}^B=(\varepsilon+l\varpi)
		\phi_{j}^A\\
	&	\left[-i\partial_x+i(k_y+A_B)\right]\phi_{j}^A=(\varepsilon+l\varpi)\phi_{j}^B
	\end{align}
and then by mapping one equation into the other, we get a second order differential equation, similar to the one used in \cite{doublelaser,doubletemps}. This gives us the spinors for $j=2,4$
\begin{widetext}
\begin{align}
&\psi_{2}(x,y,t)=\sum_{m,l=-\infty}^{\infty} e^{i k_{y} y}\left[b_{2l} \begin{pmatrix}
	1\\\xi_{1l}
\end{pmatrix}e^{i q_l (x-x_1)}+c_{2l} \begin{pmatrix}
1\\ -\frac{1}{\xi_{1l}}
\end{pmatrix}e^{-i q_l (x-x_1)}\right]\delta_{m,l}e^{-iv_F(\varepsilon+m\varpi) t}\\
&\psi_{4}(x,y,t)=\sum_{m,l=-\infty}^{\infty} e^{i k_{y} y}\left[b_{4l}
\begin{pmatrix}
	1\\\xi_{2l}
\end{pmatrix}
e^{i q_l (x-x_3)}+c_{4l}
\begin{pmatrix}
	1\\ -\frac{1}{\xi_{2l}}
\end{pmatrix}e^{-i q_l (x-x_3)}\right]\delta_{m,l}e^{-iv_F(\varepsilon+m\varpi) t}\\
& \xi_{1l}=\frac{q_l+i (k_y+A_1)}{\varepsilon +l\varpi}\\
& \xi_{2l}=\frac{q_l+i (k_y+A_2)}{\varepsilon +l\varpi}
\end{align}
\end{widetext}
corresponding to the energies 
\begin{align} 
	\varepsilon + l \varpi = s_l \sqrt{q_l^2 + (k_y + A_i)^2}
	\end{align} 
where $A_i=B_i l_B$ ($i=1,2$) and $s_l= \text{sign}(\varepsilon+l\varpi)$. As can be clearly seen, this leads to the propagating modes $q_l$ along the $x$-direction.

In region 3, where a time-oscillating laser field is applied, we use Floquet theory to derive the eigenspinors. This theory is ideal for systems with periodic time-dependent Hamiltonians, as in \eqref{Ham}. The solution of the eigenvalue equation can then be expressed in terms of Floquet modes, leading to two coupled equations
 	\begin{align}
 	\left[-i\partial_x-i(k_y-m\varpi)\right]\phi_{j}^B=(\varepsilon+m\varpi)\phi_{j}^A\\
 	\left[-i\partial_x+i(k_y-m\varpi)\right]\phi_{j}^A=(\varepsilon+m\varpi)\phi_{j}^B
 \end{align}
 which allow us to determine the spinor in region 3 as
\begin{widetext}
 \begin{align}
 &	\psi_3(x,y,t)=e^{ik_yy}\sum_{m,l=\infty}^{\infty}\left[b_{3l}\begin{pmatrix}
 		1\\
 		\Gamma_l
 	\end{pmatrix}e^{iQ_l(x-x_2)}+c_{3l}\begin{pmatrix}
 		1\\
 		-\frac{1}{\Gamma_l}
 	\end{pmatrix}e^{-iQ_l(x-x_2)}\right]J_{m-l}(\alpha_2)e^{-iv_F(\varepsilon+m\varpi)t}
 \\
 	 &\Gamma_l=s_l\frac{Q_l+i(k_y-l\varpi)}{\sqrt{(Q_l)^2+(k_y-l\varpi)^2}}\\
 	 \end{align}
 	\end{widetext}
 	including the position $x_2=d_1$ and the following propagating modes 
 	 \begin{equation}	 	
 	 Q_l=\sqrt{(\varepsilon+l\varpi)^2-(k_y-l\varpi)^2}.
 	 \end{equation}
 	 
Next, we can calculate the transmission coefficients as a function of the incident energy, magnetic field, and laser field intensities. From these we derive the transmission probabilities. We also show that the conductance can be calculated by integration over the allowed modes, in accordance with the Landauer-Büttiker formalism \cite{butker}.

\section{Transmissions and conductance}\label{cond}

Using the transfer matrix method, we connect the eigenspinors in different regions of the system. This ensures that the spinor components remain continuous at the interfaces. We assume that the barrier is infinite along the $y$-direction and therefore, the edge effects of the barriers are neglected \cite{infini1,infini2}. The method helps to analyze the behavior of the system and to calculate transmission probabilities and conductance. Indeed, we have 
\begin{align}
&	\psi_1(0)=\psi_2(0)\label{2525}\\
	& \psi_2(d_1)=\psi_3(d_1)\\
	& \psi_3(d_1+L)=\psi_4(d_1+L)\\
	& \psi_4(d_1+L+d_2)=\psi_5(d_1+L+d_2).\label{2828}
\end{align}
In our model, the magnetic barriers are represented as idealized $\delta$-function profiles. At each barrier, the spinor wavefunctions remain continuous, while their derivatives exhibit a finite discontinuity proportional to the barrier strength. This is consistent with standard treatments of sharp magnetic interfaces \cite{mag3}, and it is fully accounted for in the boundary conditions applied in (\ref{2525}-\ref{2828}). Such a formulation captures the essential physics of abrupt magnetic changes while allowing an analytic treatment of electron transmission.
These matching conditions at each interface between regions  will allow us to derive the transmission coefficients for both the propagating and evanescent modes. Taking into account the orthogonalization term $e^{iv_Fm\varpi t}$, we obtain the following eight equations
 \begin{align}
 	&\delta_{m,0}+r_m=b_{2m}+c_{2m}\label{2929}\\
 	&\delta_{m,0}\gamma_m-r_m\gamma^*_m=b_{2m}\xi_m-c_{2m}\xi_m^*\\
 	&b_{2m}e^{iq_md_1}+c_{2m}e^{-iq_md_1}=\sum_{l=-\infty}^{\infty}\left[b_{3l}+c_{3l}\right]J_{m-l}\\
 	&b_{2m}\xi_me^{iq_md_1}-c_{2m}\xi_m^*e^{-iq_md_1}=\sum_{l=-\infty}^{\infty}\left[b_{3l}\Gamma_l-c_{3l}\Gamma^*_l\right]J_{m-l}\\
 	&b_{4m}+c_{4m}=\sum_{l=-\infty}^{\infty}\left[b_{3l}e^{iQ_lL}+c_{3l}e^{iQ_lL}\right]J_{m-l}\\
 	&b_{4m} \xi_m-c_{4m}\xi_m^*=\sum_{l=-\infty}^{\infty}\left[b_{3l}\Gamma_le^{iQ_lL}-c_{3l}\Gamma^*_le^{iQ_lL}\right]J_{m-l}\\
 	&b_{4m}e^{iq_m^l d_2}+c_{4m}e^{-iq_m^l d_2}=t_m+\mathbb{0}_m\\
 	&b_{4m}\xi_m e^{iq_m d_2}-c_{4m}\xi^*_m e^{-iq_m d_2}=t_m \gamma_m-	\mathbb{0}_m \gamma_m^*\label{3636}
 \end{align}
 where we set $J_{m-l}= J_{m-l}(\alpha)$.
 To simplify the resolution of the above set, it is convenient to use the matrix approach. Here, a matrix $\mathbb{M}$ connects region $1$ to region $5$ as follows
 \begin{align}\label{Tmat}
 	\binom{\delta_{m,0}}{r_m}
=\mathbb{M}\begin{pmatrix}
 		t_m\\
 		\mathbb{0}_m
 	\end{pmatrix}
 \end{align}
such that $\mathbb{M}$ is given by
\begin{align}\label{36}
\mathbb{M}&=\mathbb{M}(1,2)\cdot\mathbb{M}(2,3)\cdot\mathbb{M}(3,4)\cdot\mathbb{M}(4,5)\\
&= 	\begin{pmatrix}
	\mathbb{M}_{11}&	\mathbb{M}_{12}\\
	\mathbb{M}_{21}&	\mathbb{M}_{22}
\end{pmatrix}\label{37}
\end{align}
where $\mathbb{M}(j,j+1)$ is the transfer matrix coupling the region $j$ with the region $j+1$, $\mathbb{0}_m$ is the zero vector, and the matrices have the forms 
 	\begin{align}
 		&\mathbb{M}(1,2)=\begin{pmatrix}
 			\mathbb{I}&\mathbb{I}\\
 			\mathbb{N}_1^+&\mathbb{N}_1^-
 		\end{pmatrix}^{-1}\cdot \begin{pmatrix}
 			\mathbb{I}&\mathbb{I}\\
 			\mathbb{N}_2^+&\mathbb{N}_2^-
 		\end{pmatrix}\\
 		&\mathbb{M}(2,3)=\begin{pmatrix}
 		\mathbb{E}^+&\mathbb{E}^-\\
 		\mathbb{K}^+&\mathbb{K}^-
 		\end{pmatrix}^{-1}\cdot \begin{pmatrix}
 			\mathbb{J}^+&\mathbb{J}^-\\
 			\mathbb{Q}^+&\mathbb{Q}^-
 		\end{pmatrix}\\
 		&\mathbb{M}(3,4)=\begin{pmatrix}
 			\mathbb{Z}^+&\mathbb{Z}^-\\
 			\mathbb{R}^+&\mathbb{R}-
 		\end{pmatrix}^{-1}\cdot \begin{pmatrix}
 			\mathbb{I}&\mathbb{I}\\
 			\mathbb{N}_2^+&\mathbb{N}_2^-
 		\end{pmatrix}\\
 		&\mathbb{M}(4,5)=\begin{pmatrix}
 			\mathbb{S}^+&\mathbb{S}^-\\
 			\mathbb{C}^+&\mathbb{C}^-
 		\end{pmatrix}^{-1}\cdot\begin{pmatrix}
 			\mathbb{I}&\mathbb{I}\\
 			\mathbb{N}_1^+&\mathbb{N}_1^-
 		\end{pmatrix}
 	\end{align}
having the following  elements 
	\begin{align}
 		&(\mathbb{N}^\pm_{1})_{m,l}=\pm(\gamma_m)^{\pm 1}\delta_{m,l}\\
 		&(\mathbb{N}^\pm_{2})_{m,l}=\pm(\xi)_m^{\pm 1}\delta_{m,l}\\
 	    &(\mathbb{E}^\pm)_{m,l}=e^{\pm iq_x^ld_1}\delta_{m,l}\\
 	     &(\mathbb{K}^\pm)_{m,l}=\pm \gamma_l^\pm e^{\pm iq_x^ld_1}\delta_{m,l}\\
 		&(\mathbb{J})_{m,l}=J_{m-l}(\alpha)\\
 		&(\mathbb{Q}^\pm)_{m,l}=\pm \Gamma^{\pm 1}_l J_{m-l}(\alpha)\\
 		&(\mathbb{Z}^\pm)_{m,l}=J_{m-l}(\alpha)e^{\pm i Q_x^l L}\\
 		&(\mathbb{R}^\pm)_{m,l}=\pm \Gamma^{\pm 1}_l J_{m-l}(\alpha)e^{\pm i Q_x^l L} \\
  		&(\mathbb{S}^\pm)_{m,l}=e^{\pm iq_x^ld_2}\delta_{m,l}\\
 		&(\mathbb{C}^\pm)_{m,l}=\pm \gamma_l^\pm e^{\pm iq_x^ld_2}\delta_{m,l}.
 	\end{align}
From \eqref{Tmat} and \eqref{37}, we can easily identify the transmission coefficients associated with different modes as a function of the system parameters
 	\begin{equation}
 		t_m=\mathbb{M}_{1,1}^{-1}\cdot\delta_{m,0}.
 	\end{equation}
 		The transfer matrix $\mathbb{M}$ is of infinite order. To solve this problem, we consider a finite number of terms from $-N$ to $N$, where $N$ satisfies the following condition: $N>\alpha$ \cite{timepot,timepot2}. 
 	Thus, the series can be expressed as follows
 	\begin{align}
 		t_{-N+k} = \mathbb{M}^{-1}_{11}[k+1, N+1], \quad k = 0, 1, \cdots, 2N.
 	\end{align}
 	Note that  the truncation condition \(N>\alpha\) is adopted as a convergence requirement ensuring that all relevant Floquet sideband contributions are retained. We tested the stability of the transmission and conductance by increasing the mode cutoff and confirmed that the key features, including resonance positions and conductance saturation, remain unchanged, indicating robustness against the cutoff parameter.

 	At this stage, we introduce the current density to determine the transmission $T_l$ and reflection $R_l$ probabilities. Using the continuity equation, we obtain the incident, transmitted, and reflected current densities as 
 	\begin{eqnarray}
 		J_{i}^0&=&v_F(\gamma_0+\gamma^*_0)\\
 		J_{t}^l&=&v_Ft^*_lt_l(\gamma_l+\gamma^*_l)\\
 		J_{r}^l&=&v_Fr^*_lr_l(\gamma_l+\gamma^*_l).
 	\end{eqnarray}
 Consequently, from the definition we obtain
 	\begin{align}
 	&	T_l=\frac{|J_{t}^l|}{|J_{i}^0|}=|t_l|^2\\	& R_l=\frac{|J_{r}^l|}{|J_{i}^0|}=|r_l|^2.
 	\end{align}
 	The total transmission probability is obtained by summing the contributions of all individual modes. Then, we have
 	\begin{equation}
 		T=\sum_{l=-\infty}^\infty T_l.
 	\end{equation}	
 The complexity of studying all transmission modes  requires a focus on the initial modes. Indeed, 
 	The resulting system consists of eight coupled equations (\ref{2929}-\ref{3636}), each admitting an infinite number of modes. In principle, solving the full system is possible, but it leads to a large $n \times n$ matrix that is difficult to handle numerically. To simplify the analysis, we restrict the study to the central band and the first sidebands ($l = 0, \pm 1, \pm 2$), resulting in a system of 40 equations.
 These are transmission $t_{0} $ without photon exchange corresponds to the central energy band $\varepsilon$, while the first $t_{\pm1} $ and second $t_{\pm2} $ sidebands correspond to $\varepsilon\pm \varpi$ and $\varepsilon\pm2 \varpi$, respectively. They are 
 	\begin{align}
 	&	t_{0}=\mathbb{M}_{1,1}^{-1}[3,3]\\
 	&	t_{-2}=\mathbb{M}_{1,1}^{-1}[1,3]\\ &t_{-1}=\mathbb{M}_{1,1}^{-1}[2,3]\\ &t_{1}=\mathbb{M}_{1,1}^{-1}[4,3]\\ &t_{2}=\mathbb{M}_{1,1}^{-1}[5,3].
 	\end{align}

At zero temperature, the conductance is defined as the average flux of fermions over half of the Fermi surface \cite{butker, conduct1}. Alternatively, it can be described as the integral of the total transmission coefficient \(T\) over \(k_y\), which is given by
\begin{equation}
G = \frac{G_0}{2\pi} \int_{-k_y^{\text{max}}}^{k_y^{\text{max}}} T(E, k_y) \, dk_y	
\end{equation}
where \(G_0\) is the unit of conductance, and \(k_y^{\text{max}}\) is the maximum wave vector component along the \(y\)-direction. The connection between the transverse wave vector \(k_y\) and the incident angle \(\phi\) is used to express \(G\) as 
\begin{equation}
G = \frac{G_0}{2\pi} \int_{-\phi^{\text{max}}}^{\phi^{\text{max}}} T(E, \phi) \cos\phi \, d\phi_0	
\end{equation}
and \(\phi_0^{\text{max}}\) can be obtained from the relation \(k_y^{\text{max}} = k \sin(\phi_0^{\text{max}})\), with \(k\) given by \eqref{E8}. This relation provides a connection between the energy \(E\) and the maximum angle \(\phi_0^{\text{max}}\) for the allowed transmission channels.

In order to explore and highlight the fundamental characteristics of our system, we perform a detailed numerical analysis of the transport properties. This analysis focuses on the behavior of the transmission channels, which are crucial for understanding the conductance of the system. By studying the contribution of each mode and its corresponding energy band, we aim to reveal how the system responds to different external parameters such as magnetic field strengths, laser intensity, and barrier distances. Our results provide insight into the resonant transmission phenomena and their effect on the total conductance, shedding light on the underlying mechanisms that govern transport in the presence of such intricate interactions.

\section{Numerical results}\label{TP}
In this section, we present a numerical analysis of our results to explore how changes in key parameters affect transmission and conductance in the system. More specifically, we analyze these quantities as a function of incident energy, barrier distance, incident angle, magnetic field, and laser field. These analyses help to clarify the role of each parameter in shaping the electron transport properties. By systematically studying these dependencies, we gain a deeper understanding of the interplay between external fields and structural configurations, which is essential for optimizing the design and performance of graphene-based devices.

\begin{figure}[ht]
	\centering
	\subfloat[]{\centering\includegraphics[scale=0.6]{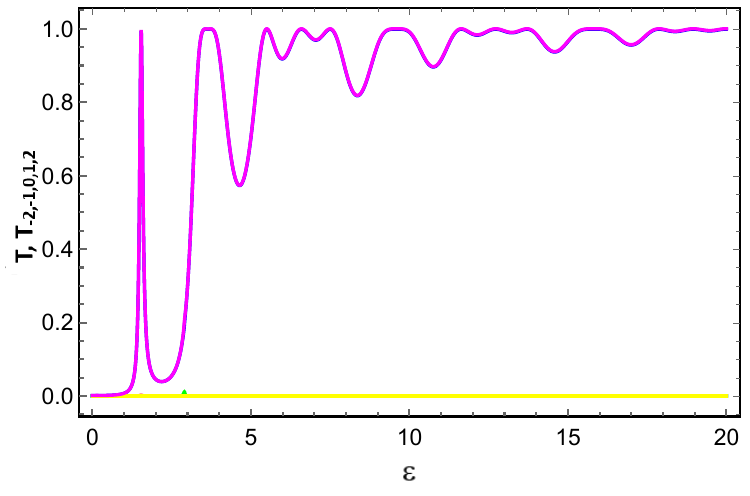}\label{fig0a}}\\
	\subfloat[]{\centering\includegraphics[scale=0.6]{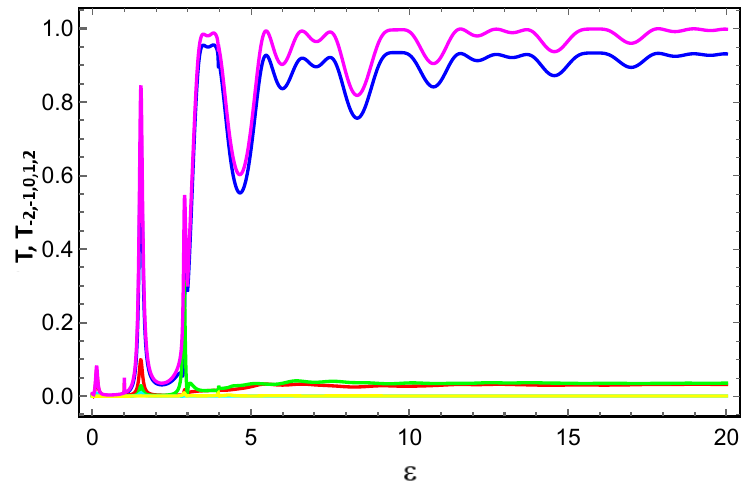}\label{fig0b}} \\
	\subfloat[]{\centering\includegraphics[scale=0.6]{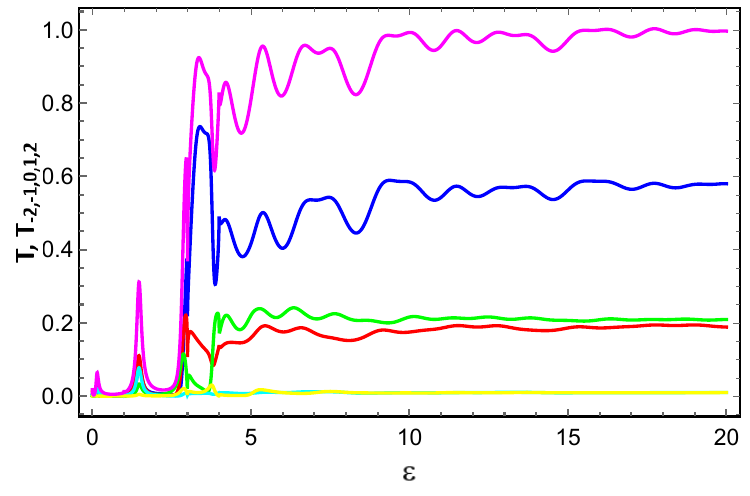}\label{fig0c}}  
	\caption{The transmissions as a function of incident energy $\varepsilon$ for normal incident $\phi=0$,  $\varpi=1$, $B_1=B_2=1$ and $d_1=d_2=L=1$. Three values of $\alpha$ (a): $\alpha=0.1$, (b): $\alpha =0.75$, (c): $\alpha=2$. Here $T$ (magenta line), $T_0$ (blue line), $T_1$ (red line), $T_{-1}$ (cyan line), $T_2$ (green line), and $T_{-2}$ (yellow line).}\label{fig0}
\end{figure}

Fig. \ref{fig0} shows the total transmission and the first five transmission modes as a function of the normal incidence energy for \(B_1 = B_2 = 1\), \(d_1 = d_2 = L = 1\), \(\varpi = \frac{\omega}{v_F}=1\) and different values of the parameter characterizing the laser field \(\alpha= \frac{E_0}{\varpi^2}\). Fig. \ref{fig0a} is plotted for very low laser irradiation \(\alpha = 0.1\). We observe that the total transmission oscillates, with the transmission without photon exchange equal to the total transmission, while the transmissions with photon exchange are almost negligible due to the low laser irradiation. This behavior is attributed to the temporal oscillation of the laser field, which generates multiple transmission modes. The amplitude of the irradiation influences the lateral modes, as discussed in our previous research \cite{Elaitouni2023, doublelaser, doublelaser+magn, Elaitouni2023A}.
We observe Fano-type resonances induced by non-adiabatic pumping from a laser field \cite{Zhu2015,W1999}. The laser field can generate quasi-bound states through photon-assisted processes between electrons and the electromagnetic field \cite{Laserquasi}. In this scenario, tunneling transport occurs via the interference between these quasi-bound states and the electronic continuum.
 These resonances diminish in the low-energy region, a phenomenon also observed for double magnetic barriers \cite{doublemag}.
In Fig. \ref{fig0b}, with \(\alpha = 0.75\), we see the appearance of the transmission by photon exchange, but the transmission without photon exchange remains dominant, almost equal to the total transmission. In Fig. \ref{fig0c}, where we increase the laser irradiation intensity to \(\alpha = 2\), we observe that transmission with photon absorption (green and red lines) becomes more significant than transmission with photon emission. Almost half of the fermions cross the barrier by photon exchange. 
We see that increasing the laser intensity increases transmission through the lateral bands but decreases transmission through the central band. This also reduces the total transmission, which suppresses the perfect transmission \cite{klien1}. Laser irradiation traps fermions in quantized energy levels \cite{Stark}, affecting transmission, and heats the graphene \cite{heat}, further delaying transmission.

\begin{figure}[ht]
	\centering
	\subfloat[]{\centering\includegraphics[scale=0.6]{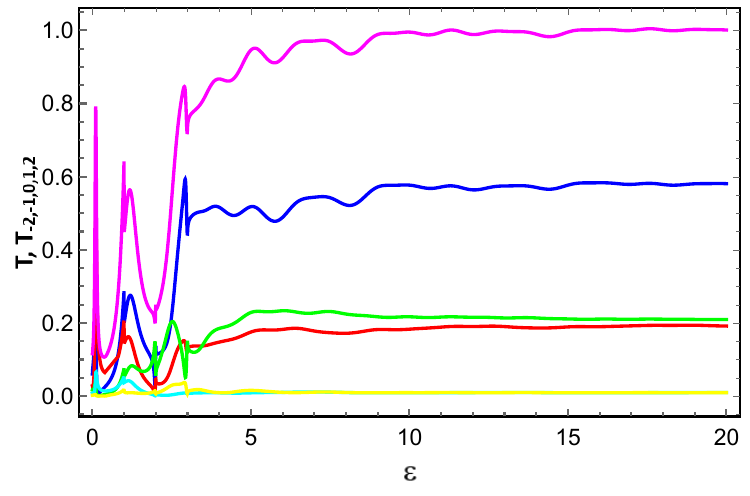}\label{fig1a}}\\
	\subfloat[]{\centering\includegraphics[scale=0.6]{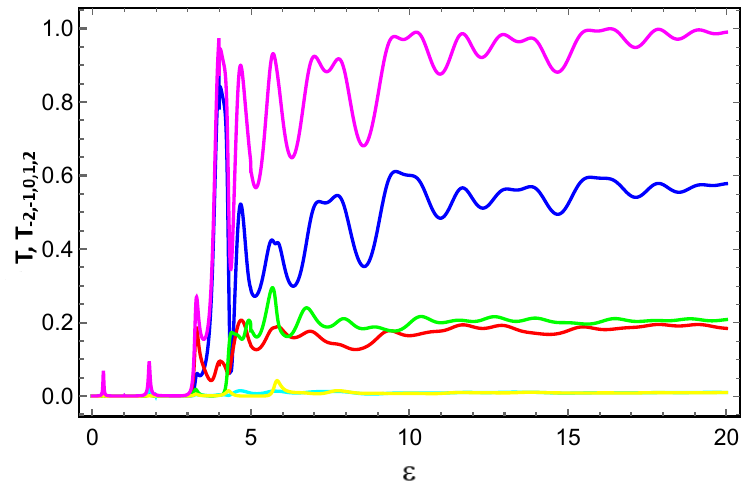}\label{fig1b}} \\
	\subfloat[]{\centering\includegraphics[scale=0.6]{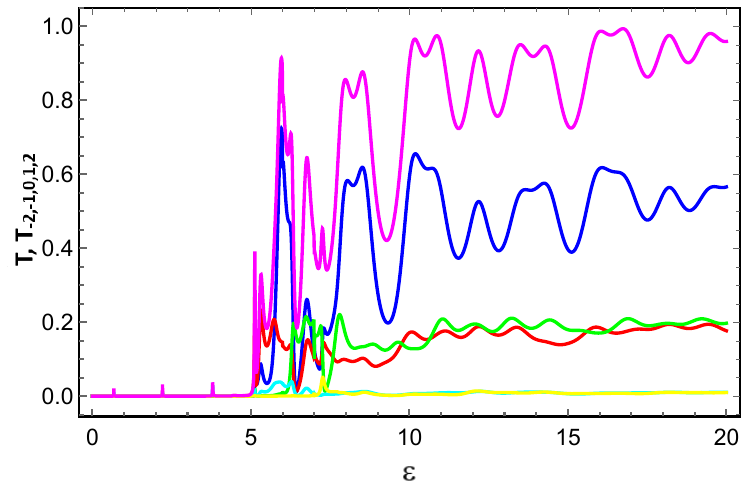}\label{fig1c}}  
	\caption{The transmissions as a function of incident energy $\varepsilon$ for normal incident,  $\varpi=1$, $\alpha=2$, $L=0.5$ and $d_1=d_2=1$. for three  values of $B$ (a): $B=1$, (b): $B=3$, (c): $B=5$. Here  $T$ (magenta line), $T_0$ (blue line), $T_1$ (red line), $T_{-1}$ (cyan line), $T_2$ (green line), and $T_{-2}$ (yellow line).}\label{fig1}
\end{figure}

{Fig. \ref{fig1} depicts the transmission as a function of incident energy $\varepsilon$ for \(\alpha = 2\), \(\varpi = 1\), \(L = 0.5\), and \(d_1 = d_2 = 1\), under varying magnetic fields (\(B_1 = B_2 = B\)). In Fig. \ref{fig1a}, for a low magnetic field \(B = 1\), weaker Fano resonances are observed, indicating that the magnetic field has a limited effect on modulating transmission at this intensity. As the magnetic field increases, additional transmission oscillations appear, as shown in Figs. \ref{fig1b} and \ref{fig1c}. At higher magnetic fields, the transmission exhibits more pronounced Fano resonances, highlighting the complex interplay between the magnetic field and the incident energy. In particular, a threshold magnetic field is required for transmission to occur, effectively acting as an effective mass \cite{masse}. This threshold suppresses transmission at low energies while enhancing the whole resonance structure. As a result, increasing the magnetic field can cause complex interactions between bound and quasi-bound states. This leads to oscillations in the energy levels, seen as peaks. These interactions can modify the transmission and confinement properties of electrons.}

\begin{figure}[ht]
	\centering
	\subfloat[]{\centering\includegraphics[scale=0.6]{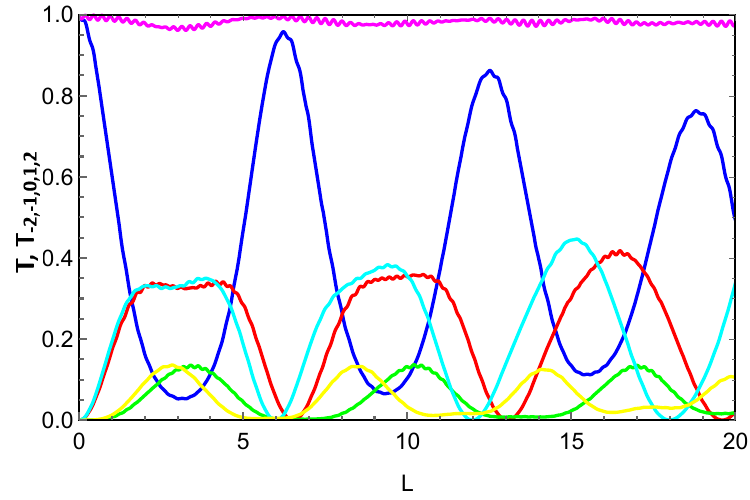}\label{fig2a}}\\
	\subfloat[]{\centering\includegraphics[scale=0.6]{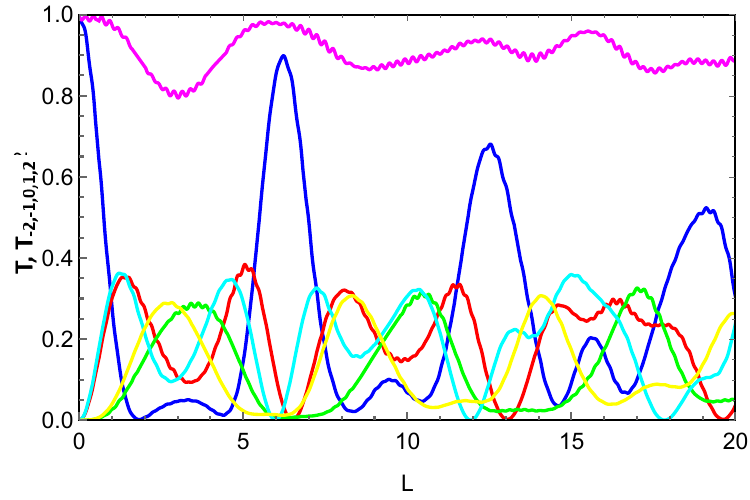}\label{fig2b}} \\
	\subfloat[]{\centering\includegraphics[scale=0.6]{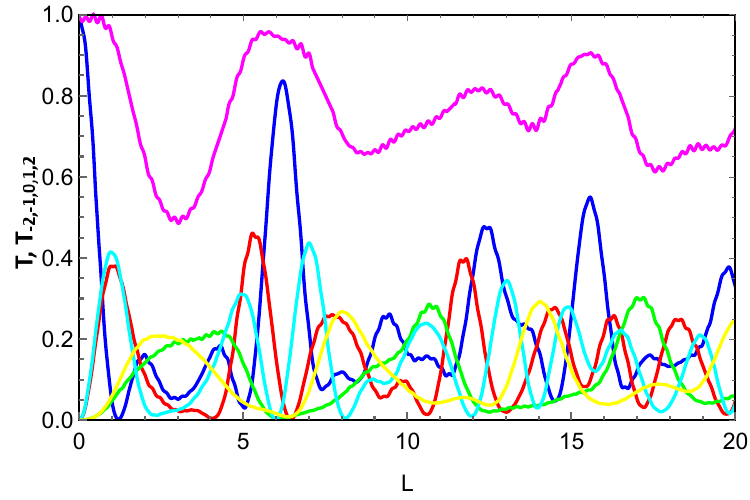}\label{fig2c}}
	\caption{(The transmissions as a function of the inter-barrier distance $L$ 
		for normal incident, $d_1=d_2=1${, $\varepsilon=15$} and $\varpi=1$. 		
		Three  values of $\alpha$ (a): $\alpha=1$, (b): $\alpha=1.5$, (c): $\alpha=2$. Here $T$ (magenta line), $T_0$ (blue line), $T_1$ (red line), $T_{-1}$ (cyan line), $T_2$ (green line), and $T_{-2}$ (yellow line).}\label{fig2}
\end{figure}

Fig. \ref{fig2} shows the transmission as a function of the inter-barrier distance \(L\),  where the laser field is applied, for different values of \(\alpha\) with \(d_1 = d_2 = 1\). 
We observe that as  long as \(L\) increases, the probability of photon diffusion increases. As a result, the interaction between the fermions and the laser field becomes stronger. This interaction leads to a reduction in transmission.
Fig. \ref{fig2a} is plotted for \(\alpha = 1\), where the total transmission oscillates around unity, indicating that a periodic perfect transmission  is observed. The transmission without photon exchange also oscillates and dominates the other transmission modes. Transmission with photon absorption is more likely than transmission with photon emission. Single-photon exchange transmission is almost equal to the probability of two-photon exchange transmission. 
As \(\alpha\) is increased, transmissions with photon emission and absorption become more probable than transmissions without photon exchange, leading to a decrease in total transmission, as shown in Fig. \ref{fig2b}. As mentioned earlier, increasing the laser field reduces the transmission without photon exchange and activates the transmission with photon exchange \cite{Elaitouni2023,Elaitouni2023A}. For Fig. \ref{fig2c} with \(\alpha = 2\), the decrease in total transmission becomes more pronounced, and transmissions with and without photon exchange begin to compete. The transmission with single or double photon exchange exceeds 40\% for certain values of \(L\). We emphasize that the total transmission is affected by the laser field intensity. In addition, transmission with photon absorption is more likely than transmission with photon emission.

\begin{figure}[ht]
	\centering
	\subfloat[]{\centering\includegraphics[scale=0.6]{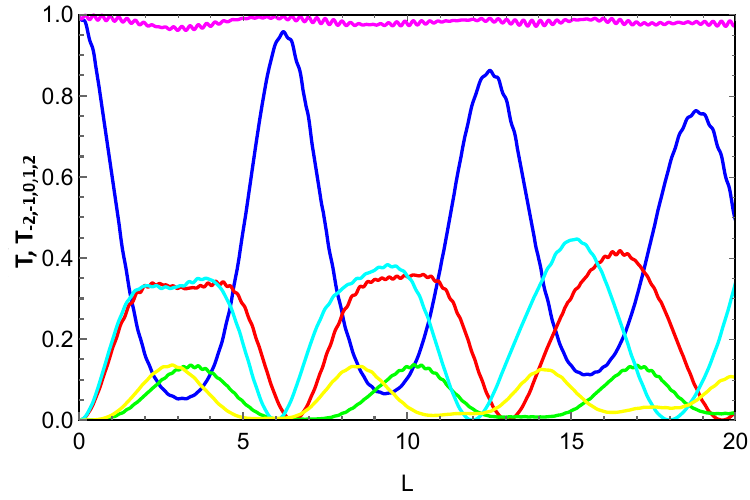}\label{fig3a}}\\
	\subfloat[]{\centering\includegraphics[scale=0.6]{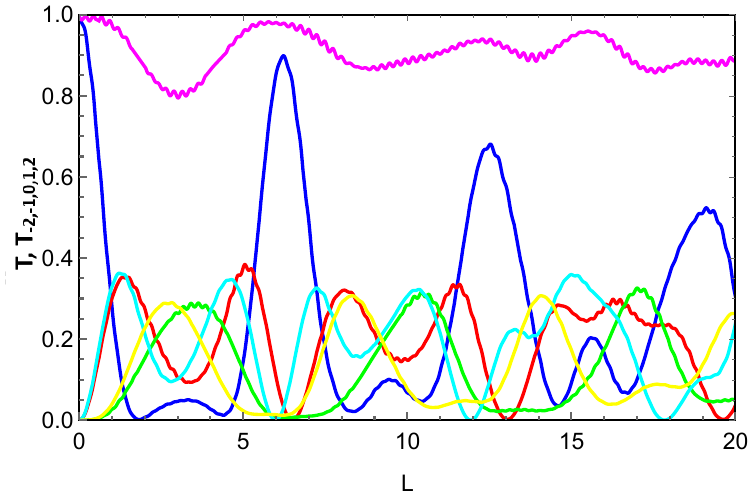}\label{fig3b}} \\
	\subfloat[]{\centering\includegraphics[scale=0.6]{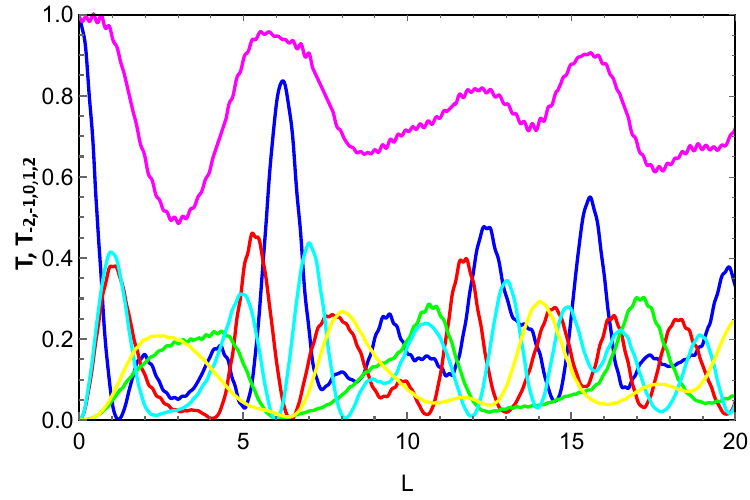}\label{fig3c}}
	\caption{ The transmissions as a function of the magnetic field ratio $B_2/B_1$  for normal incident,  $\varepsilon=15$, {$\varpi$=1} and $d_1=d_2=d=1$. Three values of $\alpha$ (a): $\alpha=1$, (b): $\alpha=1.5$, (c): $\alpha=2$. Here $T$ (magenta line), $T_0$ (blue line), $T_1$ (red line), $T_{-1}$ (cyan line), $T_2$ (green line), and $T_{-2}$ (yellow line).}\label{fig3}
\end{figure}

\begin{figure}[ht]
	\centering
	\subfloat[]{\centering\includegraphics[scale=0.6]{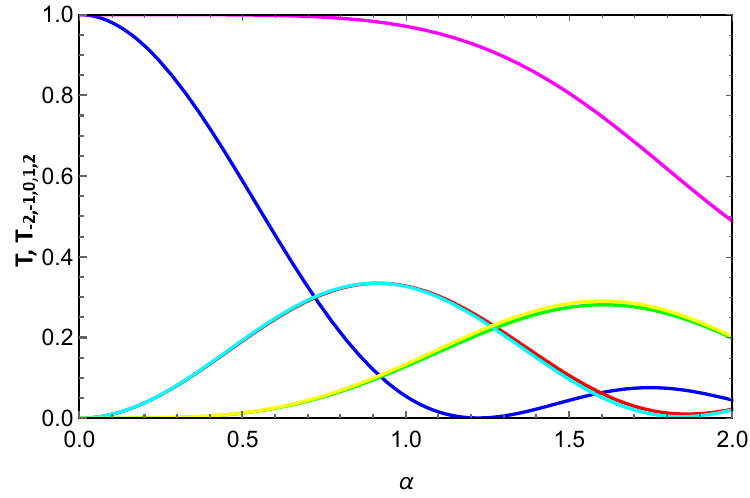}\label{fig5a}}\\
	\subfloat[]{\centering\includegraphics[scale=0.6]{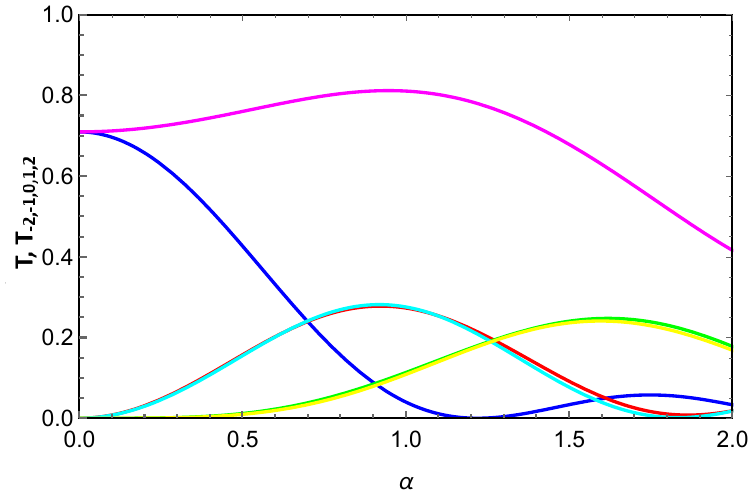}\label{fig5b}} \\
	\subfloat[]{\centering\includegraphics[scale=0.6]{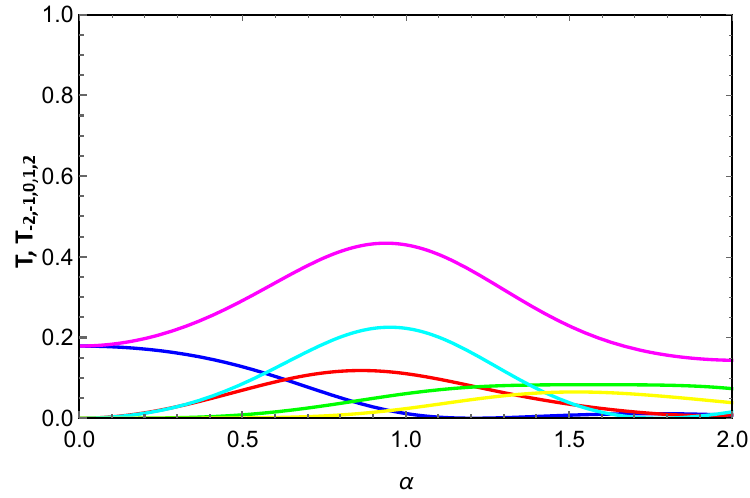}\label{fig5c}}
	\caption{The transmissions as a function of the laser field intensity $\alpha$  for normal incident,  $\varepsilon=15$, $d_1=d_2=L=1$ and $\varpi=1$. Three values of $B$ (a): $B=0.1$, (b): $B=5$, (c): $B=10$. Here  $T$ (magenta line), $T_0$ (blue line), $T_1$ (red line), $T_{-1}$ (cyan line), $T_2$ (green line), and $T_{-2}$ (yellow line).}\label{fig5}
\end{figure}

In Fig. \ref{fig3}, we plot the transmissions as a function of the magnetic field ratio \(B_2/B_1\) for different values of \(\alpha= \frac{E_0}{\varpi^2}\). We observe that the transmissions decrease as \(B_2/B_1\) increases. For \(\alpha = 1\) in Fig. \ref{fig3a}, we see that the transmissions decrease rapidly and become zero at the value \(B_2 = 8B_1\). It can be clearly seen that the transmission is more likely with the exchange of a single photon than with the exchange of two photons. 
In Fig. \ref{fig3b} with \(\alpha = 1.5\), the zero transmission point shifts to \(B_2 = 9B_1\), highlighting the effect of laser irradiation.
The combination of magnetic field and laser irradiation perturbs the fermion energy levels, reducing transmission. In addition, transmission is more likely to occur with the exchange of one photon than with the exchange of two photons. 
As \(\alpha\) increases to \(2\) in Fig. \ref{fig3c}, the same trend continues. Transmission involving the exchange of a single photon remains more likely than transmission involving the exchange of two photons. In addition, the zero transmission point continues to shift, now exceeding \(B_2 = 10B_1\). From these observations, we can conclude that increasing the magnetic field strength significantly suppresses transmission across all bands. This suppression becomes so dominant that the transmission modes are largely unaffected by further increases in laser field intensity. In other words, the magnetic field plays a critical role in blocking transmission, overshadowing the influence of the laser field at higher intensities. This highlights the interplay between the magnetic and laser fields in controlling electron transport \cite{Zhu2015,W1999}.

Fig. \ref{fig5} shows the transmission as a function of the parameter \(\alpha = \frac{E_0}{\varpi^2}\), which characterizes the laser field, for different magnetic field values \(B_2 = B_1 = B\). In Fig. \ref{fig5a}, where the magnetic field is weaker (\(B = 0.1\)), transmission without photon exchange dominates for \(\alpha < 1\). However, as \(\alpha\) increases, transmission with photon exchange becomes more likely. In addition, transmission with a single photon exchange is nearly equal to transmission with two photons, with a shift in \(\alpha\). For \(B = 5\) in Fig. \ref{fig5b}, all transmissions decrease but retain the same shape as in Fig. \ref{fig5a}. In Fig. \ref{fig5c}, where the magnetic field is stronger (\(B = 10\)), the total transmission amplitude does not exceed 50\%. Furthermore, transmission without photon exchange drops to zero at \(\alpha = 1\), and for \(\alpha > 1.5\), transmission with exchange of two photons becomes more likely than transmission with a single photon. In summary, increasing both the magnetic field intensity and the laser irradiation intensity suppresses the fermion tunneling effect. This occurs because the fields block fermions at degenerate energy levels, leading to a reduction in total transmission. The interplay between the magnetic and laser fields plays a critical role in controlling the transmission behavior. As a result, the application of stronger fields significantly limits electron transport.

\begin{figure}[ht]
	\centering
	\subfloat[]{\centering\includegraphics[scale=0.6]{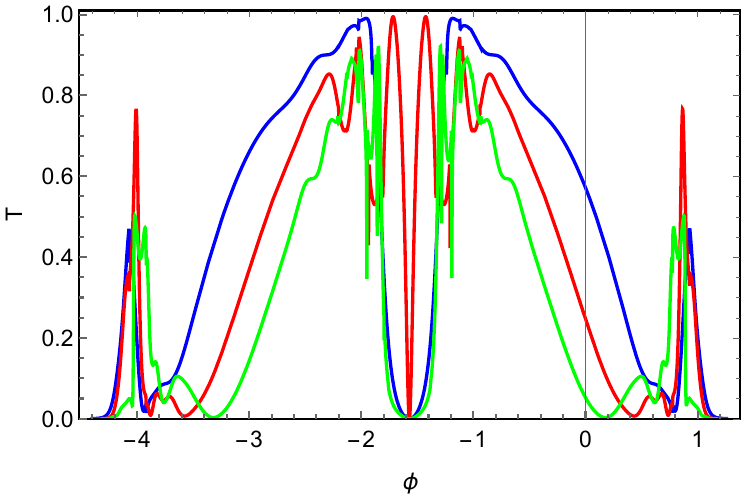}\label{fig4a}}\\
	\subfloat[]{\centering\includegraphics[scale=0.6]{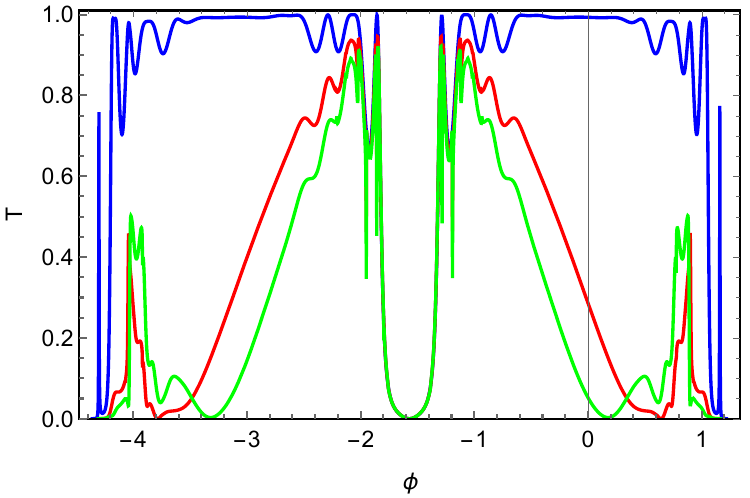}\label{fig4b}} \\
	\subfloat[]{\centering\includegraphics[scale=0.6]{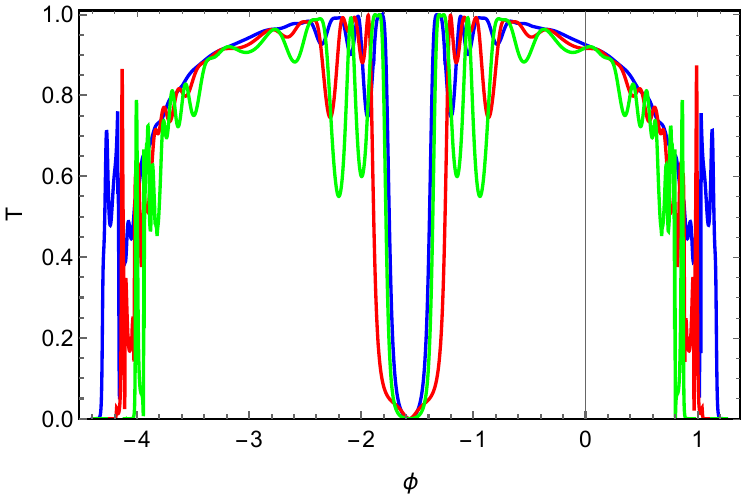}\label{fig4c}}
	\caption{The total transmission as a function of incident angle $\phi$ for $\varpi=1$ and $\varepsilon=10$. (a): $B_1=B_2=1$, $\alpha=2$ and $d_1=d_2=L=d$ for $d=0.5$ (blue line), $d=0.75$ (red line) and $d=1$ (green line). (b): $d_1=d_2=L=1$, $B_1=B_2=1$ $\alpha=0.1$ (blue line), $\alpha=1$ (red line) and $\alpha=2$ (green line). (c): $d_1=d_2=1$, $L=0.5$, $\alpha=0.75$ for $B=1$ (blue line), $B=2$ (red line) and $B=3$ (green line). }\label{fig4}
\end{figure}

Fig. \ref{fig4} presents the total transmission as a function of the incident angle \(\phi\) for various physical parameters. In Fig. \ref{fig4a}, the total transmission is plotted for three different values of the inter-barrier distance \(L\). For all three values of \(L\), the transmission drops to zero at \(\phi = -1.6\), demonstrating symmetry. As \(L\) increases, fermion has more space to diffuse and interact with the laser field, leading to a reduction in transmission. In Fig. \ref{fig4b}, the transmission is shown for three values of \(\alpha\). For \(\alpha = 0.1\) (green line), the transmission remains nearly perfect except for a narrow interval around \(\phi\) where it drops to zero. As the laser field intensity increases (red and blue curves), the transmission decreases. This reduction is due to the lower probability of coincidence between fermion and hole states \cite{Zhu2015,W1999}. Fig. \ref{fig4c} displays the transmission for three values of the magnetic field \(B\). A symmetry is observed, with transmission dropping to zero at a specific value of \(\phi\). Increasing \(B\) slightly reduces the total transmission. Additionally, the transmission consistently drops to zero at a certain \(\phi\), indicating an effect referred to as anti-resonances (perfect reflection).

r-barrier distance.

\begin{figure}[ht]
	\centering
	\subfloat[]{\centering\includegraphics[scale=0.6]{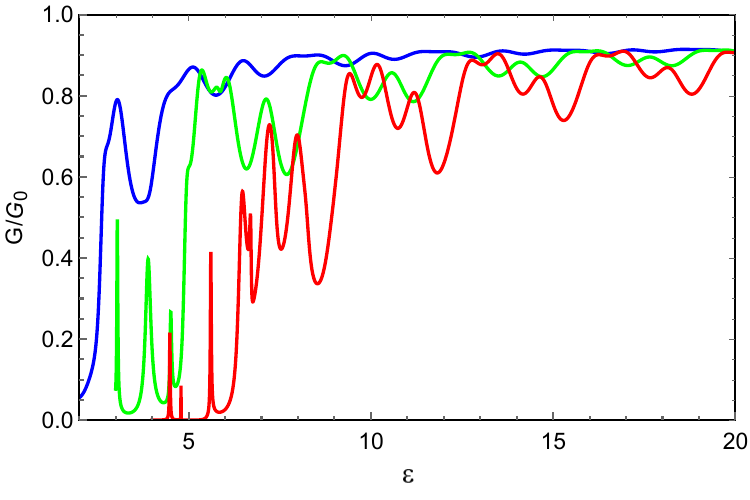}\label{fig6a}}\\
	\subfloat[]{\centering\includegraphics[scale=0.6]{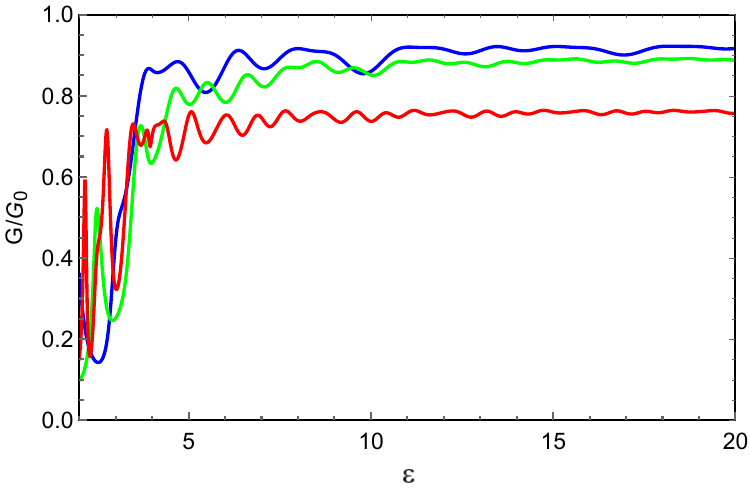}\label{fig6b}}\\
	\subfloat[]{\centering\includegraphics[scale=0.6]{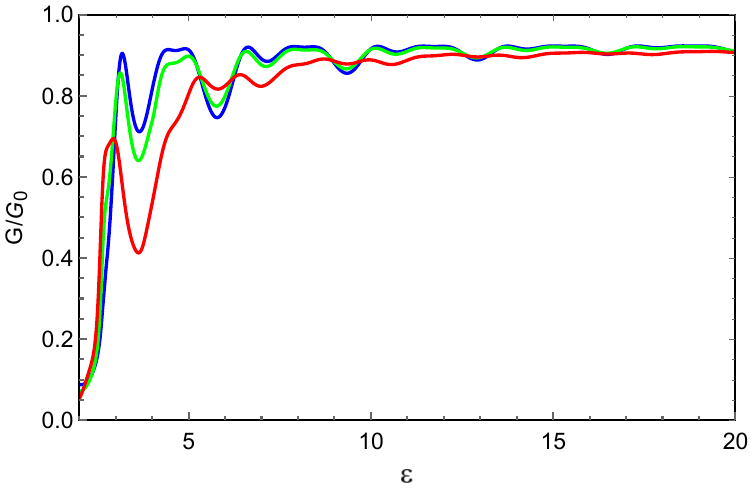}\label{fig6c}}
	\caption{The conductance as a function of the incident energy $\varepsilon$ for  $\varpi=0.5$, $B1 = B2 = B$ and $d_1=d_2=d$. (a): $d=L=1$, $B=1$ (blue line), $B=2$ (green line) and $B=3$ (red line). (b):  $d=1$, $B=1$, $L=0.5$ (blue line), $L=1.5$ (green line) and $L=3$ (red line). (c): $d=L=1$, $B=1$, $\alpha=0.5$ (blue line), $\alpha=1$ (green line) and $\alpha=2$ (red line).}\label{fig6}
\end{figure}

In Fig. \ref{fig6}, we plot the conductance at zero temperature as a function of the incident energy \(\varepsilon\) for various conditions of the physical parameters. Indeed, for three magnetic fields (\(B_2 = B_1 = B\)) in Fig. \ref{fig6a}, we observe that at low incident energies the conductance is zero, indicating that low-energy fermions cannot pass through the barrier. As the magnetic field strength increases, the conductance decreases because fewer particles can traverse the region under the influence of the applied fields. For three inter-barrier distances $L$ in Fig. \ref{fig6b}, our results show that increasing $L$ leads to a significant decrease in conductance. This means that increasing this distance gives the fermions a greater chance for destructive interference and also increases their interaction with the applied laser field. This interaction can change the energy levels of the fermions and trap them between these energy levels, resulting in a decrease in conductance.
For three values of the laser field intensities denoted by $\alpha$ in Fig. \ref{fig6c}, we observe that the increase of $\alpha$ results in a decrease of the conductance. This decrease is due to the effect of the interaction between the fermions and the laser field. As \(\alpha\) increases, the energy states of the fermions become more perturbed. This can change their trajectories and reduce their chances of crossing the barrier. The interaction with the laser field changes the energy levels available to the fermions and can introduce quantum transitions between these levels. These transitions can temporarily trap the fermions in certain quasi-bound states, reducing their mobility and ability to cross the barrier. In addition, the laser field can also induce quantum resonance phenomena in which fermions absorb and emit energy coherently with the field. These resonances can lead to destructive interferences that affect the transmission of the fermions. 
 At finite temperature, the amplitude of conductance oscillations decreases. The broadening of the Fermi distribution leads to a wider occupation of electronic states, thereby reducing the sharpness of these oscillations \cite{tempfini}.
In summary, higher magnetic fields, stronger laser fields, and larger inter-barrier distances reduce fermion transmission, leading to lower conductance.

The numerical results presented above reveal how the interplay between magnetic barriers and laser-induced photon-assisted channels governs transmission, conductance, and resonant features. They provide a foundation for drawing conclusions about the unique transport behavior and potential applications of this hybrid graphene system.

\section{Conclusion}\label{CC}

We have studied the quantum transport properties of a system consisting of two magnetic barriers, modeled by $\delta$ functions, separated by a region subjected to monochromatic linearly polarized laser irradiation. The periodic nature of the Hamiltonian of the system in time required the application of the Floquet approximation to obtain the wave function in each region. The continuity of the wave function across the five regions resulted in a system of eight coupled equations, each of which contains an infinite number of Floquet modes due to the degeneracy of the energy spectrum. To deal with the complexity of these equations, we used a matrix formalism which, together with current density calculations, allowed us to determine the transmission properties. Furthermore, the conductance was calculated using the Büttiker formula, which is commonly used in mesoscopic transport theory.
The degeneracy of the energy spectrum revealed two different transmission processes. The first involves fermions tunneling through the magnetic barriers without exchanging photons with the laser field, while the second involves photon exchange, either by absorption or emission, during transmission. 
The hybrid configuration studied here introduces transport phenomena that cannot occur in isolated magnetic or laser barriers. In our system, electrons interact sequentially with magnetic and laser fields. First, the magnetic barrier filters carriers via momentum selection. Next, the laser region induces photon-assisted transitions that open additional transmission channels. Finally, the second magnetic barrier further modifies the electron energies. This sequential interplay between magnetic confinement and photon pumping produces Fano-type resonances, tunable perfect transmission, and angle-dependent anti-resonances—features that do not appear when magnetic or laser barriers are applied individually.

Due to the inherent complexity of numerically capturing transmission across an infinite number of energy bands, our analysis focused on the central energy band corresponding to $\varepsilon$ and the first two sidebands corresponding to $\varepsilon \pm \varpi$ and $\varepsilon \pm 2\varpi$, respectively. We found that transmission without photon exchange tends to dominate the process, especially at lower laser intensities. However, as the laser field intensity increases, the probability of photon-assisted transmission increases, gradually suppressing the photon-independent process. Perfect transmission of fermions at normal incidence is observed, but shifted to lower energy regions due to the presence of the laser field and magnetic barriers.  Notably, this effect can be attenuated by tuning the system parameters, in particular the laser intensity, resulting in a gradual reduction of the total transmission. On the other hand, anti-resonance (perfect reflection), which corresponds to the complete suppression of transmission for certain incident angles, is observed in smaller energy intervals and exhibits symmetry around its zero transmission point.
Furthermore, our results show that the magnetic field shapes the transmission spectrum. Indeed, it is found that a stronger magnetic field creates more resonance peaks (Fano types) and blocks low-energy fermions, thereby reducing transmission and conductance. In addition, it is found that increasing the inter-barrier distance also increases the interaction between the fermions and the laser field. As a result, both transmission and conductance are reduced. This finding is key to understanding how barrier spacing affects fermion dynamics and can help in the design of better graphene-based devices.
 The issue of normal incidence has already been observed in graphene, as demonstrated in our previous works. At normal incidence, the transmission is nearly total, which explains why we focused our study on this case. In contrast, for oblique incidence, fermions can be controlled, allowing for the adjustment of transmission properties.

	The hybrid magnetic–laser barrier configuration offers precise control over electron transmission through tunable Fano resonances and photon-assisted channels \cite{Oka2009,McIver2020}. Magnetic barriers in graphene can be realistically approximated in the sharp-interface limit, where momentum-dependent filtering dominates transport \cite{mag1}. Angle-dependent transmission and directional electron beam shaping, which form the basis of angle-selective beam splitters, have been predicted from coherent wave-matching mechanisms in graphene structures \cite{klien1,Cheianov2006}. This hybrid mechanism could be exploited in graphene-based conductance filters, angle- or energy-selective electron beam splitters, and terahertz (THz) photodetectors, where photon-assisted processes enable efficient light sensitivity in the THz range \cite{Koppens2014,Vicarelli2012}. The interplay between magnetic confinement and laser-induced pumping provides a versatile platform for controllable electronic and optoelectronic device functionalities.

 \section*{Acknowledgment}
 P.D. and D.L. acknowledge partial financial support from FONDECYT 1231020.
 
% 	\section*{Author Contributions}
% All authors contributed equally to this work.
% All authors have read and approved the published version of the
% manuscript.
% 
% 
% \section*{Data Availability Statement}
% This manuscript has no associated data
% or the data will not be deposited. [Authors’ comment: The data that
% support the findings of this study are available on request from the
% corresponding author].
% 
% \section*{Declarations}
% Conflict of interest The authors declare that they have no known competing financial interests or personal relationships that might appear to
% influence the work presented in this paper.

\end{document}